\documentclass[prb,twocolumn,preprintnumbers,amsmath,amssymb]{revtex4}

\usepackage{amsmath,latexsym,epsf,epsfig}

\newcommand{\II}{\mbox{${\mathbb I}$}}

\newcommand{\RR}{\mbox{${\mathbb R}$}}

\newcommand{\rd}{{\rm d}}

\newcommand{\D}{{\cal D}}

\newcommand{\ph}{\varphi}
\newcommand{\phd}{\widetilde{\varphi}} 
\newcommand{\lambdad}{\widetilde{\lambda}}
\newcommand{\tx}{\widetilde{x}} 
\newcommand{\phl}{\varphi_{i,L}}
\newcommand{\phr}{\varphi_{i,R}}
\newcommand{\phz}{\varphi_{i,Z}}

\def\a{\alpha}
\def\G{\mathbb G}

\def\S{\mathbb S}
\def\mD{\mathbb D}
\def\A{\mathcal A}
\def\O{\mathcal O}
\def\mO{\mathbb O}
\def\der{\partial }

\def\ri{{\rm i}}
\def\xt{{\widetilde x}}

\def\e{{\rm e}}
\def\eps{\varepsilon}
\def\k{\kappa}
\def\v{{\bf v}}

\begin{document}
\title{Quantum wire junctions breaking time reversal invariance}
\author{Brando Bellazzini$^1$, Mihail Mintchev$^2$ and Paul Sorba$^3$}
\affiliation{
${}^1$ Institute for High Energy Phenomenology
Newman Laboratory of Elementary Particle Physics,
Cornell University, Ithaca, NY 14853, USA\\
${}^2$ Istituto Nazionale di Fisica Nucleare and Dipartimento di Fisica dell'Universit\`a di Pisa,\\
Largo Pontecorvo 3, 56127 Pisa, Italy\\ 
${}^3$ Laboratoire de Physique Th\'eorique d'Annecy-le-Vieux, 
UMR5108, Universit\'e de Savoie, CNRS,\\  
9, Chemin de Bellevue, BP 110, F-74941 Annecy-le-Vieux Cedex, France}


\begin{abstract}

We explore the possibility to break time reversal invariance at the junction of quantum wires. 
The universal features in the bulk of the wires are described by the anyon Luttinger liquid. 
A simple necessary and sufficient condition for the breaking of time reversal invariance 
is formulated in terms of the scattering matrix at the junction. 
The phase diagram of a junction with generic number of wires is investigated 
in this framework. We give an explicit classification of those critical points which can be 
reached by bosonization and study the interplay between their stability and symmetry content.

\end{abstract}

\maketitle

\section{Introduction} 

Time reversal symmetry is a fascinating subject. In this paper we 
investigate the behavior of junctions of quantum wires under time reversal 
transformations. Quantum wire networks with junctions, which attract recently 
much attention\cite{kf-92,SS,nfll-99,sdm-01,mw-02,y-02,lrs-02,cte-02,ppil-03,coa-03,rs-04,kd-05,klvf-05,gs-05,
emabms-05,ff-05,drs-06,Bellazzini:2006jb,Bellazzini:2006kh,Bellazzini:2008mn,
hc-08,drs-08,dr-08,hkc-08,adrs-08,Bellazzini:2008cs,dr-09,Bellazzini:2008fu,Ines}, are essentially 
one-dimensional systems whose transport properties 
are affected by quantum effects. The universal features in the bulk  
are captured by the Luttinger liquid theory \cite{h-81}. The junctions represent in 
this context a kind of quantum impurities (defects), where both reflection and 
transmission can take place. This fact gives origin of a complicated phase diagram, 
which has not been yet fully understood for general boundary conditions at 
the junctions, formulated in terms of the basic fermion fields.  Focussing on the case of one junction, 
we discovered\cite{Bellazzini:2006kh} in the framework of bosonization a large 
class of boundary conditions, which preserve the exact solvability of the 
Tomonaga-Luttinger (TL) model describing the Luttinger liquid in the bulk. At criticality 
these boundary conditions simply express the splitting of the electric current in the 
junction and are therefore quadratic in the fermion fields. We classified and studied 
in this setting all critical points which respect 
time reversal invariance. In this paper we extend our framework in order to cover 
also that part of the phase diagram, where the time reversal symmetry is broken. 
Recalling that the Tomonaga-Luttinger dynamics preserves time reversal invariance, 
the breaking can take place only at the junctions. In principle such kind of junctions can 
be realized\cite{ppil-03, coa-03,emabms-05,dr-08}  
by means of an external magnetic field and are therefore of practical interest. 

The previous theoretical investigations of the stability of the critical points and 
their behavior under time reversal have been mostly focussed 
on junctions with $n=3, 4$ wires. Applying the framework developed in 
Refs. \onlinecite{Bellazzini:2006kh}-\onlinecite{Bellazzini:2008fu}, 
we face below these problems for generic $n$. 

The paper is organized as follows. In the next section we define the bulk dynamics and 
boundary conditions at the junction. Using bosonization, we recall\cite{Bellazzini:2008fu} in section III 
the exact (anyon) solution of the model. In section IV we derive the current-current 
correlation function and extract the necessary and sufficient condition for the breaking of 
time reversal. We discuss here also the Kirchhoff's rules relative to the 
$U(1)\otimes {\widetilde U}(1)$ symmetry of the model. In section V we consider the 
conductance and describe the impact of time reversal breaking on it. The classification 
and parametrization of the critical points is done in section VI. In section VII we study 
the phase diagram, concentrating mainly on the symmetry content and stability of 
the fixed points. Section VIII is devoted to our conclusions. Some technical 
details are collected in the appendices.

\section{Bulk dynamics, symmetries and boundary conditions} 

\begin{figure}[tb]
\setlength{\unitlength}{0.9mm}
\begin{picture}(450,20)(-15,20)
\put(25.2,0.7){\makebox(20,20)[t]{$\bullet$}}
\put(28.5,1){\makebox(20,20)[t]{$V$}}
\put(42,11){\makebox(18,22)[t]{$E_1$}}
\put(33,17){\makebox(20,20)[t]{$E_2$}}
\put(9,4){\makebox(20,20)[t]{$E_i$}}
\put(15,-1.2){\makebox(20,20)[t]{$P$}}
\put(20,-0.8){\makebox(20,20)[t]{$x$}}
\put(34.5,-12){\makebox(20,20)[t]{$E_n$}}
\thicklines 
\put(35,20){\line(1,1){16}}
\put(35,20){\line(-1,0){9.3}}
\put(24.4,20){\line(-1,0){8}}
\put(35,20){\line(1,-1){13}}
\put(35,20){\line(1,3){6}}
\put(20,3){\makebox(20,20)[t]{$.$}}
\put(20.9,5){\makebox(20,20)[t]{$.$}}
\put(23.8,6.6){\makebox(20,20)[t]{$.$}}
\put(20,-4){\makebox(20,20)[t]{$.$}}
\put(21.9,-6){\makebox(20,20)[t]{$.$}}
\put(24.8,-7.3){\makebox(20,20)[t]{$.$}}
\put(46,31){\vector(1,1){0}}
\put(46,9){\vector(1,-1){0}}
\put(40,35){\vector(1,3){0}}
\put(20,20){\vector(-1,0){0}}
\put(15,0.7) {\makebox(20,20)[t]{$\circ$}}
\end{picture} 
\vskip 1truecm
\caption{ A star graph $\Gamma$ with $n$ edges modelling the junction of quantum wires.} 
\label{stargraph}
\end{figure}
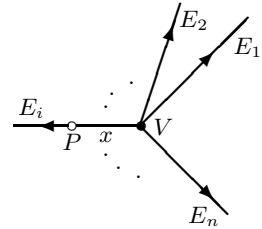

The quantum wire junction is modeled by a star graph $\Gamma$ 
of the form shown in FIG. \ref{stargraph}. 
The edges $E_i$ are half-lines and each point $P$ in the bulk of $\Gamma$ is 
uniquely determined by its coordinates $(x,i)$, where $x > 0$ is the distance to the vertex $V$ 
and $i=1,...,n$ labels the edge. $\Gamma \setminus V$ represents the {\it bulk} of the graph.  
The bulk dynamics is governed by the TL Lagrangian density 
\begin{multline}
{\cal L} = \ri \psi_1^*(\der_t - v_F\der_x)\psi_1 +  \ri \psi_2^*(\der_t + v_F\der_x)\psi_2\\ 
-g_+(\psi_1^* \psi_1+\psi_2^* \psi_2)^2 - g_-(\psi_1^* \psi_1-\psi_2^* \psi_2)^2 \, .\qquad   
\label{lagr}
\end{multline}
Here $\{\psi_\alpha (t,x,i)\,:\, \alpha =1,2\}$ 
are complex fields, $v_F$ is the Fermi velocity and 
$g_\pm \in \RR$ are the coupling constants\cite{f00}. 

The bulk theory has an obvious $U(1)\otimes {\widetilde U}(1)$ symmetry. In fact, 
the Lagrangian density (\ref{lagr}) is left invariant by the two independent 
phase transformations $(s, {\tilde s} \in \RR)$
\begin{eqnarray}
\psi_\a &\rightarrow& \e^{\ri s} \psi_\a \, , \quad \qquad \; \, 
\psi^*_\a \rightarrow \e^{-\ri s} \psi^*_\a \, ,  
\label{V}\\
\psi_\a &\rightarrow& \e^{-\ri(-1)^\a {\tilde s}} \psi_\a\,,\quad 
\psi^*_\a \rightarrow \e^{\ri(-1)^\a {\tilde s}} \psi^*_\a\, , 
\label{A}
\end{eqnarray} 
implying the current conservation laws 
\begin{equation}
\der_t \rho_\pm (t,x,i) -v_F\der_x j_\pm (t,x,i)= 0\, , 
\label{conservation}
\end{equation}
where   
\begin{equation}
\rho_\pm (t,x,i) = \left (\psi_1^*\psi_1 \pm \psi_2^*\psi_2 \right )(t,x,i) 
\label{chargedensities}
\end{equation}
are the charge densities and 
\begin{equation}
j_\pm (t,x,i) =\rho_\mp (t,x,i) 
\label{currents}
\end{equation} 
are relative currents. We adopt below also the chiral combinations 
\begin{eqnarray}
j_R(t,x,i) = \frac{1}{2}(\zeta_-j_- + \zeta_+j_+)(t,x,i)\, , 
\label{jr} \\
j_L(t,x,i) = \frac{1}{2}(\zeta_-j_- - \zeta_+j_+)(t,x,i)\, , 
\label{jl}
\end{eqnarray} 
where the real parameters $\zeta_\pm$, determined later on, are such 
that $j_L$ and $j_R$ represent the particle 
excitations moving towards and away of the vertex 
respectively. Interpreting the vertex as a 
defect, characterized by some scattering matrix, the currents $j_L$ and $j_R$ 
describe therefore the incoming and outgoing flows. 

The bulk theory is invariant also under the time reversal operation 
\begin{eqnarray} 
T \psi_1(t,x,i)T^* &=& \psi_2(-t,x,i)\, ,
\label{T1}\\
T \psi_2(t,x,i)T^* &=& \psi_1(-t,x,i)\, , 
\label{T2}
\end{eqnarray}
where $T$ is an {\it anti-unitary} operator. As already mentioned, our main goal below will be to 
investigate the impact of the vertex $V$ and the related boundary conditions on time reversal. 

The TL model (\ref{lagr}) is exactly solvable on the line $\RR$, but much care 
is needed on the graph $\Gamma$, where some boundary 
conditions must be imposed at the vertex $V$. Keeping in mind that the 
quartic bulk interactions in (\ref{lagr}) can be solved exactly 
via bosonization \cite{h-81}, it will be obviously convenient to formulate 
the boundary conditions directly in bosonic terms. In this spirit and according our  
previous comments on the chiral currents, it is quite natural to require 
that at a critical point 
\begin{equation}
j_L(t,0,i) = \sum_{k=1}^n \S_{ik}\, j_R(t,0,k) \, , \quad \forall\, t\in \RR\, . 
\label{bc1}
\end{equation} 
{}For $n=2$ this boundary condition has been first proposed and explored in Ref. \onlinecite{SS}. 
Because of scale invariance at criticality, $\S$ is a constant (momentum independent) 
unitary scattering matrix, 
\begin{equation}
\S\, \S^* = \II \, . 
\label{unit}
\end{equation} 
Since the chiral currents (\ref{jr},\ref{jl}) are Hermitian fields, one requires also 
that $\S$ has real entries, 
\begin{equation}
{\overline \S} = \S \, .  
\label{real}
\end{equation} 
Eqs. (\ref{unit}, \ref{real}) imply that $\S$ is any element of the orthogonal group $O(n)$. 
It has been shown in Refs. \onlinecite{Bellazzini:2006kh}-\onlinecite{Bellazzini:2008fu} that, 
in spite of the fact that the boundary 
condition (\ref{bc1}) is quadratic in the fields $\psi_\alpha$, it preserves the exact solvability 
of the TL model on the graph $\Gamma$. It is worth mentioning that 
this is not the case with the linear boundary conditions in 
$\psi_\alpha$, which might look at first sight simpler. 

Applying the time reversal operation (\ref{T1},\ref{T2}) to (\ref{bc1}) one infers 
\begin{equation}
j_L(t,0,i) = \sum_{k=1}^n \S^t_{ik}\, j_R(t,0,k) \, , \quad \forall\, t\in \RR\, , 
\label{bc2}
\end{equation} 
where the apex $t$ stands for transposition. Comparing (\ref{bc1}) and (\ref{bc2}) 
we conclude that {\it symmetric} scattering matrices  
\begin{equation} 
\S = \S^t  
\label{T3}
\end{equation} 
define boundary conditions which respect the time reversal invariance.  
This is the case we investigated previously in 
Refs. \onlinecite{Bellazzini:2006kh}-\onlinecite{Bellazzini:2008fu}. On the other hand, for   
\begin{equation}
\S \not= \S^t 
\label{T4}
\end{equation}
one expects breaking of time reversal. We demonstrate in Section IV 
that this is indeed the case, using the explicit form of the current-current 
correlation function. 

We conclude at this point the concise description of the bulk dynamics, symmetries 
and boundary conditions of our model and briefly describe in the next section the solution. 

\section{Solution of the TL model on a star graph} 

We look below for the solution $\psi_\alpha$ of the TL model which satisfies the 
boundary condition (\ref{bc1}) and obeys the anyon exchange relations 
\begin{multline}
\psi_\alpha^*(t,x_1,i) \psi_\alpha (t,x_2,i) =\\ 
\e^{(-1)^{\alpha} \ri \pi \k \eps(x_{12})} \psi_\alpha (t,x_2,i)\psi_\alpha^*(t,x_1,i)\, . 
\label{an1}
\end{multline} 
Here $\eps(x)$ is the sign function, $x_{12}=x_1-x_2$ and  
$\k \in \RR$ is the so called {\it statistical parameter}, which equals an even 
and an odd integer for bosons and fermions respectively. Other values of $\k$ give rise to 
Abelian anyon statistics ``interpolating'' between bosons and fermions. 

The solution on $\Gamma$ can be expressed in terms of the chiral scalar fields 
$\{\phz (\xi)\, :\, Z=L,R;\, i=1,...,n\}$, which are not independent as on the line $\RR$, but  
respect the constraints 
\begin{equation} 
\ph_{i,L} (\xi) = \sum_{j=1}^n \S_{ij} \ph_{j,R} (\xi) \, ,   
\label{bc3}
\end{equation} 
keeping track of the boundary conditions (\ref{bc1}). 
The explicit construction and a summary of the main features of $\phz$ are 
given in appendix \ref{appA}. A key point is 
the nontrivial one-body scattering matrix\cite{f0}  
\begin{equation} 
S(k) = \theta (-k) \S + \theta (k) \S^t \, , 
\label{S}
\end{equation} 
where $\theta$ is the Heaviside step function. We stress that the peculiar $k$-dependence 
of $S(k)$ respects scale invariance. 

Let us summarize now the basic features of the solution of the TL model 
with boundary conditions (\ref{bc1}). We do this essentially for two reasons. First of all 
the field $\ph$ associated with the $S$-matrix (\ref{S}) behaves quite differently 
(see appendix \ref{appA}) from its counterpart in Ref. \onlinecite{Bellazzini:2008fu}. 
Second, because we would like to keep the present paper self-contained. 

{}Following the standard bosonization procedure \cite{h-81}, we set 
\begin{eqnarray}  
\psi_1(t,x,i) &=&
z_i :\e^{\ri \sqrt {\pi} \left [\sigma \phr (vt-x) + \tau \phl (vt+x)\right ]}:\,, 
\label{psi1}\\
\psi_2(t,x,i) &=&
z_i:\e^{\ri \sqrt {\pi} \left [\tau \phr (vt-x) + \sigma \phl (vt+x)\right ]}:\,, 
\label{psi2}
\end{eqnarray}
where $: \cdots :$ denotes the normal product relative to the creation and 
annihilation operators of the fields $\phz$, namely the generators of 
the algebra (\ref{rta}). The explicit form of the normalization constants $z_i$
(including the Klein factors) is reported in appendix \ref{appA} as well. Finally, 
$\sigma$, $\tau$ and $v$ are three real parameters to be determined 
in terms of coupling constants $g_\pm$ and the statistical parameter $\k$. For this 
purpose we can assume without loss of generality that 
\begin{equation}
\sigma \geq 0\, , \qquad \sigma \not= \pm \tau  
\label{cond1}
\end{equation} 
and introduce for convenience the variables 
\begin{equation}
\zeta_\pm =\tau\pm\sigma \, . 
\label{zpm}
\end{equation} 
Plugging (\ref{psi1},\ref{psi2}) in (\ref{an1}) one gets  
\begin{equation}
\zeta_+\, \zeta_- = \kappa\, . 
\label{sys1} 
\end{equation} 
Moreover, using standard short-distance expansion for the charge densities, 
one obtains 
\begin{multline}
\rho_\pm (t,x,i) = \\
\frac{-1}{2\sqrt {\pi }\zeta_\pm } 
\left [(\der \phr)(vt-x) \pm (\der \phl)(vt+x)\right ] , 
\label{rhopm}
\end{multline} 
the normalization being fixed \cite{Bellazzini:2008fu} by the 
$U(1)\otimes {\widetilde U}(1)$-Ward identities. Inserting (\ref{psi1},\ref{psi2},\ref{rhopm}) 
in the quantum equations of motion 
\begin{multline}
\ri [\der_t +(-1)^\alpha v_F\der_x] \psi_\alpha (t,x,i)=\\
2 [g_+:\rho_+(t,x,i)\psi_\alpha : (t,x,i)\\ -(-1)^\alpha g_- :\rho_-(t,x,i)\psi_\alpha :(t,x,i)]\, , 
\label{eqm}
\end{multline}
one finds 
\begin{eqnarray} 
v \zeta_+^2 &=& v_F \k +\frac{2}{\pi}g_+ \, , 
\label{sys2}\\
v \zeta_-^2 &=& v_F \k +\frac{2}{\pi}g_- \, .  
\label{sys3}
\end{eqnarray} 
Eqs. (\ref{sys1},\ref{sys2},\ref{sys3}) provide a system for determining 
$v$ and $\zeta_\pm$ (equivalently $\sigma$ and $\tau$) in terms of 
$v_F$ and $g_\pm$. The solution is 
\begin{eqnarray}
\zeta_\pm^2 &=& |\kappa|
\left(\frac{\pi \kappa v_F+2g_+}{\pi \kappa v_F+2g_-}\right)^{\pm 1/2}\, , 
\label{z}\\ 
v&=&\frac{\sqrt{(\pi \kappa v_F+2g_-)(\pi \kappa v_F+2g_+)}}{\pi|\kappa|}\, , 
\label{v}
\end{eqnarray} 
where the positive roots are taken in the right hand side. 
The relations (\ref{z}) and (\ref{v}) represent the anyonic generalization \cite{Bellazzini:2008fu}  
of the well known result for canonical fermions ($\k=1$) in the TL 
model \cite{f1}. The conditions $2g_\pm> -\pi \kappa v_F$ 
ensure that $\sigma$, $\tau$ and $v$ are real and finite. 

Finally, in the bosonic variables $U(1)\otimes {\widetilde U}(1)$-currents $j_\pm$ 
take the form 
\begin{multline}
j_\pm (t,x,i) = \\
\frac{v}{2\sqrt {\pi }v_F \zeta_\pm} 
\left [(\der \phr)(vt-x) \mp (\der \phl)(vt+x)\right ]  
\label{jpm}
\end{multline} 
and satisfy (\ref{conservation}) by construction. 
Using (\ref{bc3}) and (\ref{jpm}) one immediately verifies that the above solution 
of the TL model on $\Gamma$ indeed satisfies the boundary condition (\ref{bc1}). 

\section{Symmetry content}

\subsection{Time reversal} 

The simplest way to investigate the behavior of the above solution under 
time reversal is to derive the two-point correlation functions of the 
currents $j_\pm$, defined by (\ref{jpm}). Using (\ref{cf1}-\ref{cf3}) one obtains 
\begin{multline} 
\langle j_+(t_1,x_1,i_1) j_+(t_2,x_2,i_2)\rangle = \\
\frac{v^2}{(2\pi \zeta_+ v_F)^2} 
[\delta_{i_1i_2} \D^2(vt_{12}-x_{12}) + \delta_{i_1i_2} \D^2(vt_{12}+x_{12}) \\
-\S_{i_1i_2} \D^2(vt_{12}+\tx_{12}) - \S^t_{i_1i_2} \D^2(vt_{12}-\tx_{12}) ] \, , 
\label{cc1}
\end{multline} 
where 
\begin{equation}
\D(\xi) = -\frac{\ri}{\xi+\ri \epsilon} 
\label{d1}
\end{equation} 
and $t_{12}=t_1-t_2,\, x_{12}=x_1-x_2$ and $\tx_{12}=x_1+x_2$. 

Let us assume for a moment that time reversal is an exact symmetry, 
or equivalently that $T$ leaves invariant the vacuum state $\Omega$. 
Then, using the anti-unitarity of $T$, one finds that 
\begin{multline} 
\langle j_+(t_1,x_1,i_1) j_+(t_2,x_2,i_2)\rangle = \\
\overline {\langle j_+(-t_1,x_1,i_1) j_+(-t_2,x_2,i_2)\rangle} 
\label{cc2} 
\end{multline} 
holds. Combining (\ref{cc1}) with (\ref{cc2}) one deduces that 
\begin{equation} 
T\Omega =\Omega \Longleftrightarrow \S = \S^t \, , 
\label{T5}
\end{equation} 
showing that the TL model on $\Gamma$ is invariant under time reversal if and only if 
$\S$ is symmetric. Otherwise, time reversal is broken, i.e. 
\begin{equation} 
T\Omega \not=\Omega \Longleftrightarrow \S \not= \S^t \, , 
\label{T6}
\end{equation} 
which confirms the conjecture made after equation (\ref{T4}) in the introduction. 
In particular, time reversal is always exact for $n=1$. For this reason we focus in what follows 
on the case $n \geq 2$.

\subsection{$U(1)\otimes {\widetilde U}(1)$-symmetry} 

Continuous symmetries on graphs are 
governed\cite{Bellazzini:2006kh,Bellazzini:2008mn} by the associated 
Kirchhoff's rules. Concerning the $U(1)\otimes {\widetilde U}(1)$-symmetry, 
using the current conservation (\ref{conservation}), one gets for the corresponding charges 
\begin{equation} 
\der_t Q_\pm = \der_t \sum_{i=1}^n \int_0^\infty \rd x \rho_\pm (t,x,i) = 
v_F \sum_{i=1}^n j_\pm (t,0,i) \, . 
\label{K1}
\end{equation} 
Inserting here (\ref{jpm}) and taking into account the boundary conditions (\ref{bc3}), one finds 
\begin{equation} 
\der_t Q_\pm =\frac{v}{2\sqrt {\pi }\zeta_\pm}\sum_{i,j=1}^n 
\left (\delta_{ij} \mp \S_{ij} \right )(\der \varphi_{j,R} )(vt) \, . 
\label{K2}
\end{equation} 
{}From this result we infer that\cite{Bellazzini:2006kh}, 
\begin{equation} 
Q_\pm - {\rm conserved} \Longleftrightarrow \sum_{i=1}^n \S_{ij} = \pm 1 \quad  \forall j=1,...,n \, . 
\label{K3}
\end{equation} 
Recalling that $Q_+$ is the electric charge and $Q_-$ the helicity of the Luttinger 
excitations, we see that only one of these quantum numbers is preserved for a generic junction\cite{f2}. 

It is worth mentioning that for junctions with $n=2$ wires the conservation of 
$Q_+$ or $Q_-$ protects the time reversal symmetry. 
Thus, junctions of three wires (T-junctions and Y-junctions) 
represent the minimal setting for breaking time reversal 
in systems preserving the electric charge $Q_+$ or the helicity $Q_-$ 
of the Luttinger liquid. Notice also that the conservation of $Q_+$ excludes the 
Dirichlet fixed point $\S= -\II$. 

\section{Conductance} 

A simple physical observable, which is sensitive to the breaking of time reversal, is the 
conductance tensor $\G_{ij}$ of the Luttinger liquid on $\Gamma$. 
In order to compute this tensor, one couples the theory 
to an external potential $A_x (t,i)$ by means of the substitution 
\begin{equation} 
\der_x \longmapsto \der_x + \ri A_x (t,i)  
\label{covder}
\end{equation} 
in eq. (\ref{lagr}). The resulting Hamiltonian is {\it time dependent} and the conductance is the coefficient 
in the linear term of the expansion of the expectation value 
$\langle J_x(t,0,i)\rangle_{A_x}$ in terms of $A_x$. For deriving $\G$ one can apply therefore 
linear response theory, which leads to\cite{Bellazzini:2006kh,Bellazzini:2008mn} 
\begin{equation} 
\G = \frac{v}{2\pi v_F \zeta_+^2} (\II - \S) \, . 
\label{G1} 
\end{equation} 
Using the condition (\ref{K3}), which ensures the conservation of the electric charge $Q_+$ 
and $\S \in O(n)$, one gets the Kirchhoff's rule for the conductance tensor 
\begin{equation}
\sum_{i=1}^n\G_{ij} = \sum_{i=1}^n\G_{ji} = 0\, , \quad  \forall j=1,...,n \, .  
\end{equation} 
If a voltage $V_i$ is applied to the edge $E_i$, the current $I_j$ flowing 
in $E_j$ is 
\begin{equation} 
I_j = \sum_{i=1}^n \G_{ji} V_i \, . 
\label{G}
\end{equation} 
Combining (\ref{T4}) with (\ref{G1}) we conclude that 
the breaking of time reversal (\ref{T4}) implies the asymmetry 
\begin{equation} 
\G \not= \G^t \, , 
\label{G2} 
\end{equation} 
a feature which has been previously observed in Refs. \onlinecite{lrs-02, coa-03, hc-08}. 
The property (\ref{G2}) 
provides an attractive experimental signature. Indeed, consider for instance the following 
two configurations with $i\not= j$. Apply first the voltage $V$ to the edge $E_i$, setting 
to 0 the voltages in all other edges, and measure the current $I_j = \G_{ji} V$. 
Repeat the same operation, applying now $V$ to the edge $E_j$ and measuring 
the current $I_i = \G_{ij} V$. If $I_i/I_j \not=1$, the system breaks time reversal. 

\bigskip 

\section{Critical points} 

\subsection{Classification} 

As already mentioned,  $\S \in O(n)$. 
There exists therefore an orthogonal matrix $\mO$, such that 
\begin{equation}
\mO \S \mO^t = \left (\begin{matrix}
\begin{matrix}r_1 & & \\ & \ddots & \\ & & r_q\end{matrix} & 0 \\
0 & \begin{matrix}\pm 1 & & \\ & \ddots & \\ & & \pm 1\end{matrix} \\
\end{matrix} \right )\, . 
\label{M1}
\end{equation} 
Here $r_i$ are $q$ rotation matrices 
\begin{equation} 
r_i = \left ( \begin{matrix}
\cos \theta_i & -\sin \theta_i \\
\sin \theta_i & \cos \theta_i \\
\end{matrix}\right )\, , 
\qquad \theta_i \in [-\pi, \pi ) \, . 
\label{M2}
\end{equation}
Let us denote by $p_\pm$ the number of eigenvalues $\pm 1$ of (\ref{M2}). Then 
$q= \frac{1}{2}(n-p_+-p_-)$ and the critical points are classified by the set 
$(p_+, p_-, \theta_1,...,\theta_q)$. From (\ref{T6}) we conclude that the time reversal 
symmetry is broken if and only if $\theta_k \not= -\pi, 0$ for some 
$k=1,...,q$. The angles $\theta_k$ thus codify the breaking of time reversal. 

\subsection{Parametrization}

$\S$ can be any element of $O(n)$, but in the physical applications 
one is mostly interested in boundary conditions which preserve the electric charge 
$Q_+$. In this case one infers from (\ref{K3}) that 
\begin{equation} 
\S \v = \v \, , \qquad \v  = \frac{1}{\sqrt n}(1,1,...,1) \, , 
\label{K4}
\end{equation}  
i.e. $\S$ leaves invariant the vector $\v$. Let $R$ be 
the orthogonal matrix (see equation (\ref{C2}) in appendix \ref{appB}), 
which rotates the vector $(0,0,...,0,1)$ in $\v$. Than $\S$ admits the 
representation 
\begin{equation} 
\S = R \left ( \begin{matrix}
\S^\prime & 0\\
0 & 1\\
\end{matrix} \right ) R^t 
\label{C1}
\end{equation} 
with $\S^\prime \in O(n-1)$. Therefore the boundary conditions which respect the 
$U(1)$-symmetry of the TL model are parametrized by the group $O(n-1)$. 
The two connected components of $O(n-1)$ give origin of two continuous families of critical points. 
Each family depends on $(n-1)(n-2)/2$ parameters, which are the angular variables parametrizing 
the elements of $O(n-1)$. 

Let us illustrate this simple general structure for $n=3$. 
The two families of critical points depend in this case on one angle  
$\vartheta \in [-\pi, \pi)$ and read 
\begin{widetext}
\begin{equation}
\S^{(1)}(\vartheta ) = 
\frac{1}{3}\left( 
\begin{array}{lll}
1+2\cos\vartheta & 1-\cos\vartheta+\sqrt{3}\sin\vartheta &  1-\cos\vartheta-\sqrt{3}\sin\vartheta\\
1-\cos\vartheta-\sqrt{3}\sin\vartheta & 1+2\cos\vartheta & 1-\cos\vartheta+\sqrt{3}\sin\vartheta\\
1-\cos\vartheta+\sqrt{3}\sin\vartheta & 1-\cos\vartheta-\sqrt{3}\sin\vartheta & 1+2\cos\vartheta
\end{array}
\right)\, ,
\label{n=31}
\end{equation} 
\begin{equation}
\S^{(2)}(\vartheta ) = 
\frac{1}{3}\left(
\begin{array}{lll}
1-2\cos\vartheta & 1+\cos\vartheta-\sqrt{3}\sin\vartheta &  1+\cos\vartheta+\sqrt{3}\sin\vartheta\\
1+\cos\vartheta-\sqrt{3}\sin\vartheta & 1+\cos\vartheta+\sqrt{3}\sin\vartheta & 1-2\cos\vartheta\\
1+\cos\vartheta+\sqrt{3}\sin\vartheta & 1-2\cos\vartheta & 1+\cos\vartheta-\sqrt{3}\sin\vartheta 
\end{array}
\right)\, ,
\label{n=32}
\end{equation} 
\end{widetext} 
confirming the recent results of Ref. \onlinecite{adrs-08}. The point 
\begin{equation} 
\S^{(1)}(-\pi ) = 
\frac{1}{3}\left( 
\begin{array}{lll}
-1&  2 & 2\\
2 &-1 & 2\\
2 &  2 & -1
\end{array}
\right)\, ,
\label{n=3c}
\end{equation} 
has been discovered by Griffith\cite{g-53} more than five decades ago in his 
pioneering work on graph models in quantum chemistry. According to (\ref{G1}), 
in this case the conductance of the Luttinger liquid is 
enhanced with respect to the line, which has been 
associated\cite{nfll-99} with the phenomenon of Andreev reflection. 
The Neumann point 
\begin{equation} 
\S^{(1)}(0) = \left( 
\begin{array}{lll}
1& 0 &0\\
0& 1& 0\\
0& 0 & 1\end{array}
\right)  
\label{n=3e}
\end{equation} 
describes instead an ideal isolator because $\G=0$. To our knowledge 
the whole family $\S^{(2)}(\vartheta )$ was 
derived\cite{f3} first in Ref. \onlinecite{Bellazzini:2006kh} and, together with the points 
(\ref{n=3c}, \ref{n=3e}), preserves time reversal invariance. Finally, the matrices $S^{(1)}(\vartheta )$ 
with $\vartheta \not=-\pi, 0$ give all critical points which violate time reversal symmetry for $n=3$. 
In a different parametrization\cite{f4} they appeared in Refs. \onlinecite{coa-03, hc-08}. 

In appendix \ref{appB} we report an explicit parametrization of the $n\times n$ 
critical $\S$-matrices. The case when the ${\widetilde U}(1)$-symmetry is 
preserved can be treated analogously\cite{Bellazzini:2008mn}.

\section{Phase diagram: boundary dimensions and stability of critical points} 

The boundary dimensions of the solution $\psi_\alpha$ capture the impact of the junction 
at criticality and can be extracted from the two-point functions
\begin{multline}
\langle \psi_1^* (t_1,x_1,i_1) \psi_1 (t_2,x_2,i_2) \rangle  = z_{i_1}z_{i_2}\\
 [{\cal D}(vt_{12}-x_{12})]^{\sigma^2\delta_{i_1i_2}} 
[{\cal D}(vt_{12}+x_{12})]^{\tau^2\delta_{i_1i_2}} \\
[{\cal D}(vt_{12}-\xt_{12})]^{\sigma \tau \S^t_{i_1i_2}} 
[{\cal D}(vt_{12}+\xt_{12})]^{\sigma \tau \S_{i_1i_2}} ,
\label{CF1} 
\end{multline}
and 
\begin{equation}
\langle \psi_2^* (t_1,x_1,i_1) \psi_2 (t_2,x_2,i_2) \rangle = 
(\ref{CF1})\quad {\rm with}\quad  \sigma \leftrightarrow \tau\, . 
\label{CF2}
\end{equation} 
Performing the scaling transformation 
\begin{equation} 
t \mapsto \varrho t\, , \quad x \mapsto \varrho x\, , \quad \varrho >0\, , 
\label{dilatation}
\end{equation}
in (\ref{CF1}, \ref{CF2}), one obtains 
\begin{multline}
\langle \psi_\alpha^* (\varrho t_1,\varrho x_1,i_1) 
\psi_\alpha (\varrho t_2,\varrho x_2,i_2) \rangle  = \\
\varrho^{-D_{i_1i_2}} \langle \psi_\alpha^* (t_1,x_1,i_1) \psi_\alpha (t_2,x_2,i_2) \rangle \, , 
\label{scaling}
\end{multline}
where 
\begin{equation}
D=(\sigma^2 +\tau^2) \II_n +\sigma \tau (\S + \S^t) \, . 
\label{D}
\end{equation} 
The eigenvalues $d_i$ of the matrix $D/2$ are the scaling dimensions. 
If time reversal is broken ($\S\not= \S^t$), some of the eigenvalues 
of $\S$ are necessarily complex. Notice however that the eigenvalues of the 
combination $\S+\S^t$ are all real and 
\begin{equation}
d_i=\frac{1}{2} (\sigma^2 + \tau^2) + \sigma \tau s_i\, , 
\label{di}
\end{equation} 
$s$ being the $n$-vector 
\begin{equation} 
s = (\cos \theta_1, \cos \theta_1, \cos \theta_2, \cos \theta_2, ... , \cos \theta_{2q}, \pm 1, ... , \pm1) \, .
\label{svector}
\end{equation} 
In appendix \ref{appC} we prove that the mixing between $\psi_1$ and $\psi_2$ 
produces vanishing additional eigenvalues and therefore 
does not affect the spectrum (\ref{di},\ref{svector}). 

Recalling that the scaling dimension on the line is \cite{cft} 
\begin{equation} 
d^{({\rm line})} = \frac{1}{2} (\sigma^2 + \tau^2) \, , 
\label{dline}
\end{equation} 
one deduces from (\ref{di}) the {\it boundary dimensions} 
\begin{equation} 
d_i^{({\rm boundary})} = \sigma \tau s_i =\frac{\xi_{+}^{2}-\xi^{2}_{-}}{4}s_{i}\, , 
\label{dboundary}
\end{equation} 
which control\cite{c-84, cft} the stability of the critical points. The direction $i$ 
at a critical point $\S$ of the phase diagram 
is stable (unstable) if $d_i > 0$ ($d_i<0$). We call the point $\S$ 
{\it completely} stable if all relative directions are stable. 
Using (\ref{sys2},\ref{sys3}), the boundary dimension $d_i$ can 
be rewritten in our case in the form 
\begin{equation}
d_i^{({\rm boundary})} = \frac{1}{2\pi v}(g_+ - g_-)s_{i}\, , \quad v>0\, ,  
\label{dboundaryTL}
\end{equation} 
where $v>0$ is given by (\ref{v}). 
It is natural to consider at this point the two regimes of {\it repulsive} ($g_{+}>g_{-}$) and 
{\it attractive} ($g_{+}<g_{-}$) anyonic interactions. From (\ref{dboundaryTL}) 
one concludes that in the repulsive case the direction $i$ is stable if 
$s_i>0$. Vice versa, in the attractive case stability requires $s_i<0$. 

The direction $i$ in the phase diagram is called flat if $d_i=0$. This happens for 
$\cos \theta_i = 0$ and/or $g_+=g_-$. The last case is very special: there is no interaction between 
$\psi_1$ and $\psi_2$ (see eq. (\ref{lagr})), all boundary dimensions vanish and all 
directions are flat. 

It is worth stressing that the above considerations 
concern the phase diagram of the system without 
symmetry constraints. According to Section IV however, 
the Kirchhoff's rules controlling the symmetry content 
of the TL model on $\Gamma$, impose such constraints.   
If one requires for instance $U(1)$-symmetry, the condition (\ref{K3}) implies 
that $s_i=1$ in at least one direction. 
Therefore, for attractive interactions with $U(1)$ symmetry there are no 
completely stable points.  The same conclusion holds in the repulsive case 
with ${\widetilde U}(1)$ symmetry. 

As already mentioned, imposing time reversal symmetry implies that 
$\theta_i=-\pi$ or $\theta_i= 0$ for all $i=1,...,q$, which severely restricts the phase diagram. 
In particular, the only completely stable fixed points are 
$\S=\II$ (for $g_+>g_-$) and $\S=-\II$ (for $g_+<g_-$), corresponding to 
Neumann and Dirichlet boundary conditions respectively. Allowing 
for breaking of time reversal leads to a richer phase diagram, which admits 
whole families of non trivial ($\S\neq\pm\II$) completely stable critical points. 

Let us consider for illustration the phase diagram for $n=3$ (Y-junction). 
We have shown above that all critical points, respecting the electric charge 
conservation, are given by (\ref{n=31}, \ref{n=32}). The corresponding eigenvalues 
are 
\begin{equation}
s^{(1)}= (\cos \theta, \cos \theta , 1)\, , \quad 
s^{(2)}= (1, 1, -1)\, , 
\end{equation} 
showing that the family $\S^{(2)}(\theta )$, which preserves time reversal symmetry, 
does not contain completely stable points. The time reversal breaking family $\S^{(1)}(\theta)$ 
contains instead the non trivial completely stable fixed points with 
$\cos \theta >0$ in the repulsive regime and $\cos \theta <0$ in the attractive one. In this sense 
complete stability is favored by time reversal breaking. 

We stress in conclusion that the above algorithm can be applied for analyzing the stability 
of the critical points under perturbation with any composite operator involving the 
basic fields $\psi_\alpha(t,x,i)$. Some examples of quadratic operators are considered in 
appendix \ref{appD}.

\section{Conclusions} 

We investigated above the behavior under time reversal of a Luttinger junction 
with any number of edges $n$ and satisfying the boundary conditions (\ref{bc1}). 
As expected, time reversal invariance can be broken by boundary effects,  
in spite of the that fact that the bulk theory preserves this symmetry. The following 
two exceptions are worth mentioning. Time reversal symmetry is always preserved for $n=1$. 
The same conclusion holds for $n=2$, provided that the electric charge $Q_+$ is conserved. 

The results of this paper give a global view on the phase diagram of the system 
with boundary conditions (\ref{bc1}) and 
the framework allows to investigate both the symmetry content and the stability of the 
critical points. It turns out that the phase diagram has two connected components, 
corresponding to those of the group $O(n)$ and therefore depending on 
$n(n-1)/2$ parameters, which describe irrelevant boundary couplings. 
In this classification the critical points, 
which respect the electric charge conservation, form a $O(n-1)$-subfamily. 
A simple criterion (\ref{T4}) allows to distinguish the points which violate 
time reversal invariance from those which preserve it. The stability 
of the critical points is controlled by the relative boundary dimensions. For generic $n$ 
we derived these dimensions in explicit form (\ref{dboundaryTL}), establishing 
their dependence on the boundary conditions and the bulk couplings. 
The analysis of the critical points, which are stable in all directions of the phase 
diagram, reveals that except of the Neumann point $\S=\II$ for repulsive 
interactions and the Dirichlet point $\S=-\II$ in the attractive case, all other 
completely stable points violate time reversal invariance. 

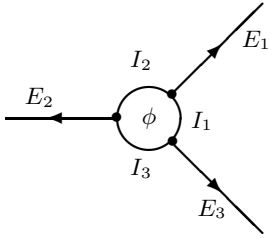
\begin{figure}[tb]
\setlength{\unitlength}{0.9mm}
\begin{picture}(450,20)(-15,20)
\put(25.2,1.6){\makebox(20,20)[t]{$\phi$}}
\put(42,11){\makebox(18,22)[t]{$E_1$}}
\put(9,3.5){\makebox(20,21)[t]{$E_2$}}
\put(34.5,-12){\makebox(20,20)[t]{$E_3$}}
\put(33,1){\makebox(20,20)[t]{$I_1$}}
\put(24,10){\makebox(20,20)[t]{$I_2$}}
\put(24,-6){\makebox(20,20)[t]{$I_3$}}
\put(20.4,0.9){\makebox(20,20)[t]{$\bullet$}}
\put(28.6,4.3){\makebox(20,20)[t]{$\bullet$}}
\put(28.7,-2.7){\makebox(20,20)[t]{$\bullet$}}
\thicklines 
\put(35.2,20){\circle{10}}
\put(39,24){\line(1,1){13}}
\put(30,20){\line(-1,0){16}}
\put(39,16){\line(1,-1){13}}
%
%
\put(46,31){\vector(1,1){0}}
\put(46,9){\vector(1,-1){0}}
\put(20,20){\vector(-1,0){0}}
%
\end{picture} 
\vskip 2truecm
\caption{ A graph with 3 external and 3 internal edges.} 
\label{graph}
\end{figure} 

As already mentioned in the introduction, the simplest realization of devises, 
violating time reversal invariance, uses\cite{ppil-03, coa-03,emabms-05,dr-08}  
magnetic fields. An example, which frequently appears in the literature \cite{coa-03, hc-08}, 
is the configuration shown in Fig. 2. One has three external half lines 
and a ring composed of three compact internal edges and three junctions. 
A magnetic flux $\phi$ is crossing the ring. The complete field theory analysis of the Luttinger liquid 
on a graph with this geometry is very complicated problem, which is beyond the 
scope of the present paper. One approximate way to face the problem could 
be to use the star product approach\cite{ks-01} or 
the ``gluing" technique\cite{Mintchev:2007qt, Ragoucy:2009hf} for 
deriving the $3\times 3$ scattering matrix relative to the {\it external} edges. 
Although a bit complicated \cite{ks-01, Ragoucy:2009hf, Schrader:2009su}, 
this $S$-matrix can be used afterwards for developing a simplified model 
with {\it one effective} junction. Clearly, such an approach does not provide 
the conductance of the internal edges $I_i$. 

The generalization of the results of this paper to off-critical junctions represents also a 
challenging open problem. The study of the rich spectrum \cite{Ines} of effects 
away of equilibrium is essential in this respect. Another interesting subject is the study of 
networks with several junctions. We are currently investigating these issues.

\acknowledgments
We thank Pasquale Calabrese for his interest in this work and for fruitful discussions. 
Correspondence with In\`es Safi is also kindly acknowledged. 
The research of B.B. has been supported in part by the NSF grant PHY-0757868. 

\appendix

\section{Chiral fields on $\Gamma$}
\label{appA}

The chiral scalar fields 
\begin{multline}
\ph_{i,R}(\xi ) = \int_{0}^{\infty} \frac{\rd k}{\pi \sqrt{2k}}
\left[a^\ast_i(k) \e^{\ri k\xi} +
a_i (k) \e^{-\ri k\xi}\right ] \, , \\ 
\ph_{i,L} (\xi) = \int_{0}^{\infty} 
\frac{\rd k}{\pi \sqrt {2k}} 
\left[a^\ast_i(-k) \e^{\ri k\xi} +
a_i (-k) \e^{-\ri k\xi}\right ] \, , \; \; \,  
\label{pirl}
\end{multline} 
are the building blocks of the solution (\ref{psi1},\ref{psi2})). 
On $\Gamma$ the generators $\{a_i(k), a^*_i(k)\}$ obey the following 
{\it deformation} 
\begin{multline} 
[a_i(k)\, ,\, a_j(p)] = [a^*_i (k)\, ,\, a^*_j (p)] = 0 \, , \\
[a_i(k)\, ,\, a^*_j (p)] = 2\pi [\delta (k-p)\delta_{ij} + \delta(k+p)S_{ij}(k)] \, , 
\label{rta}
\end{multline} 
of the standard canonical commutation relations. Here $S(k)$ is the 
one-body scattering matrix defined by (\ref{S}). Besides (\ref{rta}), 
we impose also the constraints 
\begin{eqnarray} 
a_i(k) &=& \sum_{j=1}^n S_{ij}(k) a_j(-k) \, , 
\label{c1}\\ 
a^*_i(k) &=& \sum_{j=1}^n a^*_j(-k) S_{ji}(-k) \, , 
\label{c2}
\end{eqnarray}
which are consistent, because $S(k) S(-k)=\II$, and imply (\ref{bc3}). 
Equations (\ref{rta}-\ref{c2}) define a special {\it reflection-transmission} algebra $\A$, 
which has been introduced in a more general form in the study 
\cite{Liguori:1996xr,Mintchev:2002zd,Mintchev:2003ue,Mintchev:2004jy} of point-like 
defects in integrable systems. Notice that although $k$-dependent, $S(k)$ is scale invariant. 

Time reversal is realized in the algebra $\A$ by means of 
\begin{equation}
T a_i(k) T^* = -a_i(-k)\, , \quad T a^*_i(k) T^* = -a^*_i(-k) \, .  
\label{T7}
\end{equation}
In fact, (\ref{T7}) imply 
\begin{eqnarray}
T \phr(t-x) T^* &=& -\phl (-t+x)\, , 
\label{T8}\\
T \phl (t+x)T^* &=& -\phr(-t-x) \, , 
\label{T9}
\end{eqnarray} 
which implement in turn the time reversal transformation 
(\ref{T1},\ref{T2}) on the solution (\ref{psi1},\ref{psi2}). 

For the construction of correlation functions we adopt the Fock 
representation of $\A$. We denote by $\Omega$ and $(\cdot \, ,\, \cdot )$ 
the Fock vacuum state and the scalar product, 
using for the vacuum expectation values of the operators $\O_k$ the short notation 
\begin{equation}
(\Omega \, , \O_1\cdots \O_n \Omega ) = \langle \O_1\cdots \O_m \rangle \, . 
\label{notation1}
\end{equation} 
Since $a_i(k)\Omega =0$, the basic correlators are 
\begin{multline}
\langle a_i(p)a^\ast_j(q)\rangle  = 
2\pi \left [\delta_{ij}\, \delta (p-q) 
+ S_{ij}(p)\, \delta (p+q)  \right ]\, , \\
\langle a^{\ast i}(p )a_j(q)\rangle = 0 \, ,\qquad \qquad \qquad 
\qquad \qquad \qquad \qquad \quad \,    
\label{fock2}
\end{multline} 
which imply 
\begin{eqnarray} 
\langle \varphi_{i_1,R}(\xi_1)\varphi_{i_2,R}(\xi_2)\rangle &=&  \nonumber \\
\langle \varphi_{i_1,L}(\xi_1)\varphi_{i_2,L}(\xi_2)\rangle &=& \delta_{i_1i_2} u(\mu \xi_{12})\, ,  
\label{cf1}
\end{eqnarray} 
\begin{eqnarray} 
\langle \varphi_{i_1,L}(\xi_1)\varphi_{i_2,R}(\xi_2)\rangle &=&  \S_{i_1i_2} u(\mu \xi_{12})\, , 
\label{cf2} \\
\langle \varphi_{i_1,R}(\xi_1)\varphi_{i_2,L}(\xi_2)\rangle &=& \S^t_{i_1i_2} u(\mu \xi_{12})\, ,  
\label{cf3}
\end{eqnarray} 
where $\xi_{12} = \xi_1-\xi_2$, 
\begin{equation} 
u(\xi) = -\frac{1}{\pi} \ln (\ri \xi + \epsilon)\, , \quad \epsilon >0\, , 
\label{cf4}
\end{equation} 
and $\mu>0$ is an infrared mass parameter \cite{Grignani:1988fx}. The normalization 
constants $z_i$ which occur in (\ref{psi1}, \ref{psi2}) depend 
on $\mu$ in the following way 
\begin{equation}
z_{i}=(2\pi)^{-1/2}\mu^{[(\sigma^2+\tau^2)+2\sigma\tau \S_{ii}]/2} \eta_i\, ,
\label{zi}
\end{equation}
where $\eta_i$ are the {\it anyon} Klein factors needed to ensure the correct
anyon exchange relations on different edges of the graph $\Gamma$. 
A simple representation is 
\begin{equation} 
\eta_i=:\e^{\pi\ri (\alpha_i + \alpha^*_i)}: \, , 
\label{KL1}
\end{equation}
where $\{\alpha_i,\, \alpha_i \, :\, i=1,...,n\}$ generate the auxiliary algebra 
\begin{equation} 
[\alpha_i\, ,\, \alpha_j] = [\alpha^*_i\, ,\, \alpha^*_j] = 0\, , \quad 
[\alpha_i\, ,\, \alpha^*_j] = \ri \frac{\k}{2} \epsilon_{ij} \, , 
\label{KL2}
\end{equation} 
with $\epsilon_{ij}=-1$ for $i<j$, $\epsilon_{ii}=0$ and $\epsilon_{ij}=1$ for $i>j$. 

It is worth stressing that there is an action principle behind 
the whole structure (\ref{pirl}-\ref{cf4}). The action can be written 
in terms of the combinations  
\begin{eqnarray} 
\ph_i (t,x) &=& \frac{1}{2} \left [\phr (t-x) + \phl (t+x) \right ]\, , \\
\phd_i (t,x) &=& \frac{1}{2} \left [\phr (t-x) - \phl (t+x) \right ]\,  
\label{A1}
\end{eqnarray} 
and the auxiliary fields $\{\lambda_i (t,x),\, \lambdad_i(t,x)\}$ as follows. 
The bulk and boundary actions action are 
\begin{widetext} 
\begin{equation}
{\cal S}_{\rm bulk}=\int_{-\infty}^{+\infty}\rd t\int_{0}^{+\infty}\rd x\, 
\sum_{i=1}^n \left [\lambda_i (\partial_x \ph_i + \partial_t \phd_i) + 
\lambdad_i (\partial_t \ph_i + \partial_x \phd_i)\right ](t,x) \, ,  
\label{bulka}
\end{equation} 
\begin{eqnarray}
{\cal S}_{\rm boundary}=
\frac{1}{2} \sum_{i=1}^n \int_{-\infty}^{+\infty}\rd t\, 
(\lambda_i \lambda_i - \lambdad_i \lambdad_i +
\ph_i \ph_i + \phd_i \phd_i )(t,0) +   \qquad \qquad \nonumber \\
\frac{1}{4}\sum_{i,j=1}^n \int_{-\infty}^{+\infty}\rd t\, \left [\phd_i (\S+\S^t)_{ij}\phd_j 
-\ph_i (\S+\S^t)_{ij}\ph_j -2\ph_i (\S-\S^t)_{ij}\phd_j\right ](t,0) \, ,  
\label{bounda}
\end{eqnarray} 
\end{widetext}
respectively. The total action ${\cal S} = {\cal S}_{\rm bulk}+ {\cal S}_{\rm boundary}$ is non degenerate, 
$\lambdad$ and $\lambda$ being the conjugate momenta of $\ph$ and $\phd$ respectively. 
In agreement with (\ref{T6}), the only term breaking time reversal invariance is the 
term proportional to $\S - \S^t$ in (\ref{bounda}). The bulk 
variation involves only ${\cal S}_{\rm bulk}$. Varying 
with respect to $\lambda$ and $\lambdad$, one gets the duality relations 
\begin{align}
\partial_{t}\phd(t,x,i)=&-\partial_{x}\ph(t,x,i)\,,\\ 
\partial_{x}\phd(t,x,i)=&-\partial_{t}\ph(t,x,i)\,.
\end{align} 
The bulk variation with respect to $\ph$ and $\phd$ gives analogous relations between 
$\lambda$ and $\lambdad$. The boundary variation involves both 
${\cal S}_{\rm bulk}$ and ${\cal S}_{\rm boundary}$ and, 
as easily verified, generates the boundary condition (\ref{bc3}). 

A final comment concerns an interesting interplay between locality and time reversal 
symmetry on $\Gamma$. A standard computation shows that at space-like 
separated points $t^2_{12}<x^2_{12}$ 
\begin{multline} 
[\ph (t_1,x_1,i)\, ,\, \ph(t_2,x_2,j)] = \\
-[\phd (t_1,x_1,i)\, ,\, \phd(t_2,x_2,j)]=  
\frac{\ri}{4} \left ( \S^t- \S \right )_{ij}\, , 
\label{comm1}
\end{multline} 
implying that the time reversal breaking on $\Gamma$ is accompanied by 
violation of locality of $\ph$ and $\phd$.\cite{f5} One can easily check 
however that this violation does not affect the locality of the currents $j_\pm$, 
which belong to the observables of the theory. 

More about quantum field theory on graphs (also away of criticality) can be found in Refs. 
\onlinecite{Bellazzini:2006jb}-\onlinecite{Bellazzini:2008mn}, \onlinecite{Bellazzini:2008cs} and 
\onlinecite{Mintchev:2007qt}-\onlinecite{Schrader:2009su}, where some basic 
elements\cite{ks-00, qg} of the spectral theory of differential operators on graphs 
(``quantum graphs"), have been used.

\bigskip 

\section{Critical $\S$-matrices for generic $n$}
\label{appB} 

First of all, the matrix $R$ which rotates the vector $(0,0,...,0,1)$ in ${\bf v}$ 
can be taken in the form 
\begin{equation} 
R_{ij} = \begin{cases} 
0 & \quad \text{if $i<j=1,2,...,n-1$}\, , \\
\frac{-1}{\sqrt{(n-j)(n-j+1)}} & \quad \text{if $i>j=1,...,n-1$}\, , \\
\sqrt{\frac{n-i}{n-i+1}} & \quad \text{if $i=j=1,...,n-1$}\, , \\
\frac{1}{\sqrt{n}} &  \quad \text{if $i=1,...,n,\, , j=n$}\, , \\
\end{cases}
\label{C4}
\end{equation} 
As well known, the matrix $\S^\prime \in O(n-1)$ can be parametrized in terms of the 
$(n-1)(n-2)/2$ rotation matrices $\{r_{i,j}(\vartheta_{ij})\, : i,j=1,...,n-1,\, i<j\}$ each of them 
rotating at the angle $\vartheta_{ij}$ in the $ij$-plane. 
If $\det (\S^\prime) = 1$ one has 
\begin{equation}
\S^\prime = \left ( \prod_{i=n-2}^1r_{i,n-1} \right ) 
\left (\prod_{i=n-3}^1r_{i,n-2}\right ) \cdots \left (r_{2,3} r_{1,3}\right ) r_{1,2} \, . 
\label{C2}
\end{equation} 
The only delicate point is the domain of the generalized Euler 
angles $\vartheta_{ij}$, which turns out to be \cite{HRR} 
\begin{equation} 
\vartheta_{ij} \in \begin{cases} 
[-\pi , \pi) &\quad \text{for $j=i+1$}\, , \\
[-\pi/2 , \pi/2] & \quad \text{for $j>i+1$}\, .
\end{cases}
\label{C3}
\end{equation} 
Finally, in the case $\det (\S^\prime) = -1$ one can simply multiply the 
right hand side of (\ref{C3}) by the matrix $r$, which {\it reflects} for instance along the first axis. 

\section{The $\psi_1-\psi_2$ mixing}
\label{appC} 

The mixing between $\psi_1$ and $\psi_2$ is described by the two-point functions
\begin{multline}
\langle \psi_1^* (t_1,x_1,i_1) \psi_2 (t_2,x_2,i_2) \rangle  = z_{i_1}z_{i_2}\\
 [{\cal D}(vt_{12}-x_{12})]^{\sigma \tau \delta_{i_1i_2}} 
[{\cal D}(vt_{12}+x_{12})]^{\sigma \tau \delta_{i_1i_2}} \\
[{\cal D}(vt_{12}-\xt_{12})]^{\sigma^2 \S^t_{i_1i_2}} 
[{\cal D}(vt_{12}+\xt_{12})]^{\tau^2 \S_{i_1i_2}} ,
\label{CF3} 
\end{multline}
and 
\begin{equation}
\langle \psi_2^* (t_1,x_1,i_1) \psi_1 (t_2,x_2,i_2) \rangle = 
{\rm (\ref{CF3})}\quad {\rm with}\quad  \sigma \leftrightarrow \tau\, . 
\label{CF4}
\end{equation} 
Combining (\ref{CF3},\ref{CF4}) with eqs. (\ref{CF1},\ref{CF2}), one finds that under 
a scaling transformation (\ref{dilatation}) a generic two-point function transforms according to 
\begin{multline}
\langle \psi_{\alpha_1}^* (\varrho t_1,\varrho x_1,i_1) 
\psi_{\alpha_2} (\varrho t_2,\varrho x_2,i_2) \rangle  = \\
\varrho^{-\mD_{\alpha_1\alpha_2\, ,\, i_1i_2}} \langle \psi_{\alpha_1}^* (t_1,x_1,i_1) \psi_{\alpha_2} (t_2,x_2,i_2) \rangle \, , 
\label{generalscaling}
\end{multline}
where $\mD$ is the $2n\times 2n$ matrix 
\begin{equation} 
\mD = 
\left ( \begin{matrix} 
D & B \\ 
B^t & D \\ 
\end{matrix}\right )\, , 
\label{extD}
\end{equation} 
with $D$ given by (\ref{D}) and 
\begin{equation} 
B=2 \sigma \tau \II_n +\sigma^2 \S^t + \tau^2 \S \, . 
\label{Bmatrix}
\end{equation} 
The eigenvalues the matrix $\mD/4$ 
provide the dimensions capturing the $\psi_1-\psi_2$ mixing. We will prove now that 
$n$ of the eigenvalues of $\mD/4$ vanish and that the remaining $n$ coincide precisely 
with the dimensions $d_i$ given by (\ref{di}). For this purpose we compute the 
characteristic polynomial ${\rm det}(\mD -x \II_{2n})$. First we move to the basis in which 
$\S$ has the form (\ref{M1},\ref{M2}), performing the the transformation 
\begin{equation} 
\left ( \begin{matrix} 
\mO &  0 \\ 
0 & \mO \\ 
\end{matrix}\right ) 
\left ( \begin{matrix} 
D & B \\ 
B^t & D \\ 
\end{matrix}\right ) 
\left ( \begin{matrix} 
\mO^t &  0 \\ 
0 & \mO^t \\ 
\end{matrix}\right ) = 
\left ( \begin{matrix} 
D_{\rm d} & B_{\rm bd} \\ 
B^t_{\rm bd} & D_{\rm d} \\ 
\end{matrix}\right )\, . 
\label{bdD}
\end{equation}
In this basis $D_{\rm d}$ is diagonal, whereas $B_{\rm bd}$ is block diagonal. At this point we 
use the identity\cite{Gantmacher} 
\begin{equation} 
\det \left ( \begin{matrix} 
M &  N \\ 
P & Q \\ 
\end{matrix}\right ) = \det (M) \det (Q-P M^{-1} N) \, ,  
\label{gant}
\end{equation} 
where $M, N, P$ and $Q$ are $n\times n$ blocks and $M$ is invertible. Let us  
apply (\ref{gant}) to $\det (\mD - x\II_{2n})$ with $\mD$ given by (\ref{bdD}). For 
$D_{\rm d}-x\II_n$ to be invertible we assume for the moment that 
$x\not= \sigma^2 + \tau^2 +2\sigma \tau s_i$ with $s_i$ defined by (\ref{svector}). One gets 
\begin{multline} 
\qquad \qquad \qquad \qquad \det (\mD - x\II_{2n}) = \\
\det (D_{\rm d} - x\II_n) \det (D_{\rm d} - x\II_n + B_{\rm bd} (D_{\rm d} - x\II_n)^{-1} B_{\rm bd}^t)\, .  
\label{det1}
\end{multline} 
Being determinants of diagonal and of block diagonal matrices, 
the two factors in the right hand side of (\ref{det1}) are easily computed. One finds  
\begin{eqnarray} 
\det (D_{\rm d} - x\II_n) = \prod_{i=1}^n (x-\sigma^2 - \tau^2 -2\sigma \tau s_i ) \, , 
\label{det2}\\
 \det (D_{\rm d} - x\II_n + B_{\rm bd} (D_{\rm d} - x\II_n)^{-1} B_{\rm bd}^t) = 
 \nonumber \\ 
 \frac{\prod_{i=1}^n [x(x-2\sigma^2 - 2\tau^2 -4\sigma \tau s_i )]}
 {\prod_{i=1}^n (x-\sigma^2 - \tau^2 -2\sigma \tau s_i )} \, . \qquad 
 \label{det3}
 \end{eqnarray} 
 Notice that the factor (\ref{det2}) cancels precisely the denominator 
 of (\ref{det3}). Therefore, the characteristic polynomial we are looking for is 
 \begin{equation} 
 \det(\mD - x\II_{2n}) =  
 \prod_{i=1}^n [x(x-2\sigma^2 - 2\tau^2 -4\sigma \tau s_i )] \, , 
 \label{det4}
 \end{equation}
 which extends for any $x$ by continuity and proves our statement. 
 
\section{Composite two-fermion operators}
\label{appD} 

We examine here the stability of the critical points under the perturbation with 
the composite operators 
\begin{multline}  
\Phi_1(t,x,i) = :\psi_1^*\psi_2:(t,x,i) \sim \\
:\e^{\ri \sqrt {\pi} \zeta_-\left [\phr (vt-x) - \phl (vt+x)\right ]}:\, , 
\label{D1}
\end{multline}
\begin{multline}
\Phi_2(t,x,i) = :\psi_2^*\psi_1:(t,x,i) \sim \\
:\e^{-\ri \sqrt {\pi} \zeta_-\left [\phr (vt-x) - \phl (vt+x)\right ]}:\, . 
\label{D2}
\end{multline}
The relative two-point correlation functions are easily derived. One finds  
\begin{multline}
\qquad \qquad \quad 
\langle \Phi_1^* (t_1,x_1,i_1) \Phi_1 (t_2,x_2,i_2) \rangle  = \\
\langle \Phi_2^* (t_1,x_1,i_1) \Phi_2 (t_2,x_2,i_2) \rangle  \sim \\ 
[{\cal D}(vt_{12}-x_{12})]^{\zeta_-^2 \delta_{i_1i_2}} 
[{\cal D}(vt_{12}+x_{12})]^{\zeta_-^2\delta_{i_1i_2}} \\
[{\cal D}(vt_{12}-\xt_{12})]^{-\zeta_-^2 \S^t_{i_1i_2}} 
[{\cal D}(vt_{12}+\xt_{12})]^{-\zeta_-^2 \S_{i_1i_2}} \, ,
\label{D3} 
\end{multline}
\begin{multline}
\qquad \qquad \quad 
\langle \Phi_1^* (t_1,x_1,i_1) \Phi_2 (t_2,x_2,i_2) \rangle  = \\
\langle \Phi_2^* (t_1,x_1,i_1) \Phi_1 (t_2,x_2,i_2) \rangle  \sim \\ 
[{\cal D}(vt_{12}-x_{12})]^{-\zeta_-^2 \delta_{i_1i_2}} 
[{\cal D}(vt_{12}+x_{12})]^{-\zeta_-^2\delta_{i_1i_2}} \\
[{\cal D}(vt_{12}-\xt_{12})]^{\zeta_-^2 \S^t_{i_1i_2}} 
[{\cal D}(vt_{12}+\xt_{12})]^{\zeta_-^2 \S_{i_1i_2}} \, .
\label{D4} 
\end{multline}
As before, the response of (\ref{D3}, \ref{D4}) under the scaling transformation (\ref{dilatation}) 
defines the $2n\times 2n$ matrix 
\begin{equation} 
{\widetilde \mD} = \zeta_-^2  
\left ( \begin{matrix} 
-1 &  1\\ 
1 & -1 \\ 
\end{matrix}\right ) \otimes (\S+\S^t-2\II)\, ,    
\label{D5}
\end{equation} 
the dimensions of the operators (\ref{D1}, \ref{D2}) being the eigenvalues of ${\widetilde \mD}/2$. 
One easily finds that $n$ of these eigenvalues vanish. The remaining $n$ are given by 
\begin{equation} 
{\widetilde d}_i = \zeta_-^2 (1-s_i) \, , 
\label{D6}
\end{equation} 
where $s_i$ are defined by (\ref{svector}). 
Subtracting from (\ref{D6}) the dimensions of the same operators on the line, one finds 
the nontrivial boundary dimensions 
\begin{equation} 
{\widetilde d}_i^{\, (\rm boundary)} = -\zeta_-^2 s_i  \, . 
\label{D7}
\end{equation} 
At this point one can repeat the analysis performed in section VII for perturbations with 
a single fermion operator. Comparing (\ref{dboundaryTL}) and (\ref{D7}) and using that 
$\zeta_-^2 > 0$, we see that in the attractive regime $g_+<g_-$ the stability properties 
of the critical points under the two different perturbations are the same. 
In the repulsive case $g_+>g_-$ the behavior is inverted. 
The directions which were stable become unstable under perturbations with 
(\ref{D1}, \ref{D2}) and vice versa.


\begin{thebibliography}{99}
\bibitem{f00} 
The combinations $g_{2,4}=2(g_{-}\mp g_{+})$ are often used as well.
\bibitem{f1} 
The usually used parameter $K$ in our notation
coincides for $\kappa=1$ with $\zeta_{-}^2=\zeta_+^{-2}$.
\bibitem{f4} The parameter $g_c$, adopted in Ref. \onlinecite{hc-08}, is related 
to $\vartheta$ by $\sin \vartheta = \pm \frac{2\sqrt{3} g_c}{3+g_c^2}$, the two different signs 
corresponding to the families $\chi_\pm$.
\end{thebibliography}
\end{document}